\documentclass[11pt]{article}
\usepackage{moriond2000,epsfig}

\def\Journal#1#2#3#4{{#1} {\bf #2}, #3 (#4)}

\def\be{\begin{equation}}
\def\ee{\end{equation}}
\def\bea{\begin{eqnarray}}
\def\eea{\end{eqnarray}}

\def\simless{\mathbin{\lower 3pt\hbox
   {$\rlap{\raise 5pt\hbox{$\char'074$}}\mathchar"7218$}}} 
\def\simgreat{\mathbin{\lower 3pt\hbox
   {$\rlap{\raise 5pt\hbox{$\char'076$}}\mathchar"7218$}}} 

\newcommand{\fmmm}[1]{\mbox{$#1$}}
\newcommand{\scnd}{\mbox{\fmmm{''}\hskip-0.3em .}}
\newcommand{\scnp}{\mbox{\fmmm{''}}}
\begin{document}
\vspace*{4cm}
\title{A SPECTROSCOPIC SURVEY FOR STRONG GALAXY--GALAXY LENSES}

\author{ J.P. WILLIS$^{1,2}$, P.C. HEWETT$^1$, S.J. WARREN$^3$, G.F. LEWIS$^4$}

\address{1. Institute of Astronomy, Madingley Road, Cambridge CB3 0HA,
UK. \\ 2. Departamento de Astronom{\'{\i}}a y Astrof{\'{\i}}sica,
P. Universidad Cat{\'{o}}lica, Avenida Vicu{\~{n}}a Mackenna 4860,
Casilla 306, Santiago 22, Chile. (present address) \\ 3. Blackett Laboratory,
Imperial College of Science Technology and Medicine, Prince Consort
Road, London SW7 2BZ, UK. \\ 4. Anglo-Australian Observatory, P.O. Box
296, Epping, NSW 1710, Australia.}

\maketitle\abstracts{We present a spectroscopic survey for strong
galaxy--galaxy lenses. Exploiting optimal sight--lines to massive,
bulge--dominated galaxies at redshifts $z \sim 0.4$ with wide--field,
multifibre spectroscopy, we anticipate the detection of 10--20 lensed
Lyman--$\alpha$ emitting galaxies at redshifts $z \simgreat 3$ from a
sample of 2000 deflectors. Initial spectroscopic observations are
described and the prospects for constraining the emission--line
luminosity function of the Lyman--$\alpha$ emitting population are
outlined.}

\section{Introduction}

Despite considerable advances over the past two decades in the study
of individual gravitational lens systems, the assembly of large,
uniformly--selected samples of systems multiply imaged by individual
galaxies has proved extremely hard. Searches for individual examples
of strong lensing have relied on the examination of a sample of
objects, such as quasars or flat--spectrum radio--sources, where a
large fraction of the sample lie at high redshift. Thus, towards each
object there is a significant path--length over which an intervening
deflector may interpose itself close to the line--of--sight. The lens
search proceeds through the identification of sources whose
morphology, multiple images or extended arcs for example, is
consistent with the effects of lensing. Further imaging, at different
wavelengths, and spectroscopy is then necessary to establish the
source as a bona--fide lens and to obtain redshifts for the source and
the deflecting galaxy. In practice, obtaining the redshifts is very
difficult, particularly for radio--selected objects, and in the
compilation of Kochanek et al (1999: {\it
http://cfa-www.harvard.edu/castles}) only 19 of the 45 lensed systems
possess both deflector and source redshifts.

An alternative search strategy is to examine optimal lines--of--sight
by identifying a population of very effective deflectors, where it is
known that any source lying behind the deflector will be significantly
lensed, and then to examine the spectra of the deflectors for evidence
of lensed background sources. Miralda--Escud\'{e} and
Leh\'{a}r \cite{mira92} pointed out that provided the surface density
of faint, small, galaxies at high redshift is large, significant
numbers of galaxy--galaxy lenses should exist. Subsequent
observational developments have shown that the surface density of
high--redshift, star--forming objects is indeed
large \cite{steid96,hu98}.  Provided a suitable sample of deflectors
can be identified the optimal line--of--sight search strategy offers
significant advantages, including i) high efficiency, the probability
a lens will be seen along a line--of--sight is significant, ii) the
deflector and source redshifts may be readily acquired, allowing the
full lensing geometry to be defined, iii) the small, but extended,
star--forming objects lead to resolved gravitational lenses, not
unlike the radio--rings arising from morphologically extended radio
emission, which provide much greater constraints on the deflector
masses than the more familiar two-- or four--image lenses of
unresolved quasars.

Using APM measures of United Kingdom Schmidt Telescope $B_JRI$ plates
it is possible to identify the ideal population of deflectors
\---\ massive, bulge--dominated, galaxies at redshift $z\sim 0.4$,
essentially half-way between ourselves and any high redshift source.
Specifically, locating the population of relatively bright, $m_R\le
20$, red, $B_J-R \ge 2.2$, galaxies with redshifts $0.25 \le z \le 0.6$
is straightforward \cite{war93}.  The galaxy population has a
surface density of $\sim 50\,$deg$^{-2}$ and associated with each
galaxy there is an area of sky, $\sim 1\,$arcsec$^2$, in which any
distant source will be multiply imaged, with an associated increase in
brightness of a factor $\simgreat 10$. These early--type galaxies represent
essentially optimal lines--of--sight to search for examples of strong
lensing.

The presence of a lens is revealed by the detection of an anomalous
emission line in the spectrum of one of the target distant early-type
galaxies, so obtaining spectra of a large sample of the deflector
galaxies represents the first stage in the lens survey.  Examination
of intermediate--resolution optical spectra of an initial sample of
160 colour--selected early--type galaxies revealed the presence of an
emission line at $5589$\AA \ in a galaxy with redshift
$z=0.485$. Follow--up spectroscopy \cite{war98} and imaging \cite{war99}
have confirmed the B0047--2808 system as an optical Einstein ring with
the source, a star--forming galaxy at $z=3.595$, the first confirmed
example of a normal galaxy lensing another normal galaxy and a
demonstration of the viability of the optimal line--of--sight survey
strategy.

\section{Spectroscopic observations and candidate selection}

\begin{figure}
\vbox{
\centerline{
\psfig{figure=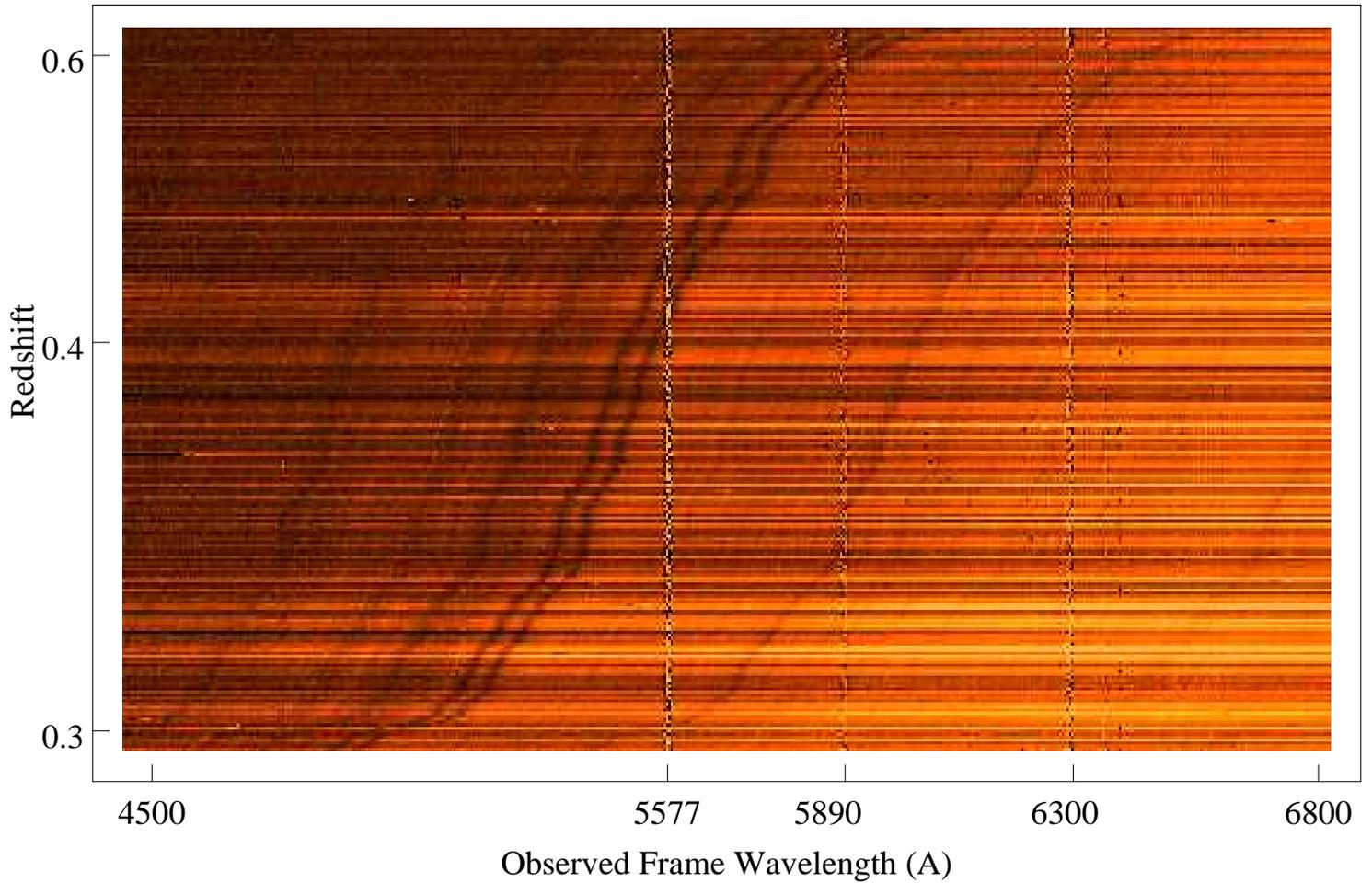,width=5.0in,angle=270.0}
}
\caption{Observed--frame spectra of 485 distant early--type
galaxies, redshifts $0.3 \le z \le 0.6$, arranged by increasing
redshift. A number of prominent night--sky features are visible as
vertical lines while features present in the galaxies, such as Calcium
H+K and the G--band move to longer wavelength with increasing redshift.}  
\label{fig-1}
}
\end{figure}

With an efficient method for acquiring spectra along many optimal
lines--of--sight there is the prospect of obtaining a large sample,
$\sim 20$ objects, of spatially resolved gravitationally lensed systems.
The low--surface density of the galaxies on the sky means the
Anglo--Australian Telescope's 2dF multifibre instrument, with a
$3\,{\rm deg}^2$ field, is ideally suited to the initial spectroscopy.
In September 1998 we obtained spectra of $\sim 500$ early--type
galaxies over two nights using the 2dF facility. Total
exposure times of $\sim 8000\,$s produce galaxy spectra for which the
completeness of redshift measurement is $95\%$ and in which anomalous
emission lines of fluxes $\sim 5\times 10^{-17}\,$erg
s$^{-1}$cm$^{-2}$ may be reliably detected i.e. fluxes comparable to
those seen in high--redshift galaxy samples \cite{hu98} can
be reached. The unlensed fluxes are a factor $\sim 10$ fainter.
Example spectra of galaxies from 2dF are shown in
Figure~\ref{fig-1}. 

Candidate gravitational lenses are identified by applying an automated
emission--line detection algorithm to the early--type galaxy
spectra. The identification software matches a template early--type
galaxy SED (derived from the mean of the sample) to each early--type
galaxy spectrum via a wavelength--dependent transformation. The
transformation is derived from the median smoothed ratio of the two
spectra. Subtraction of the transformed template from the individual galaxy
spectra removes large scale ($\lambda \simgreat 100${\AA}) continuum
variations while retaining small--scale differences such as narrow
($\lambda < 50${\AA}) emission lines. Such emission features may then
be identified using standard matched
filter techniques \cite{pratt78}. The effectiveness
of the emission--line detection routine is demonstrated by the
identification of [OII]3727 emission in 20{\%} of the early--type
galaxy sample (104/485 galaxies). Four anomalous emission lines,
consistent with gravitationally--lensed Lyman--$\alpha$ emission, have
also been identified utilising this technique. Candidate lenses must be
confirmed via a second observation of the emission--line
prior to follow--up observations to obtain the source
redshift, via observation of a second emission feature \cite{war98}, and the
morphology of the lensed emission \cite{war99}.

\section{Constraining the luminosity function of high-redshift
Lyman--$\alpha$ emitting galaxies using gravitational lensing}

A spectroscopic survey for gravitational lenses, employing a
quantitative detection algorithm and a well--defined sample of
deflectors, permits a unique experiment to probe the luminosity
function of high-redshift Lyman--$\alpha$ emitting galaxies \---\ to
fainter flux limits than currently achievable.

The line--of--sight to each deflector may be considered as magnifying
a region of the distant source plane. The total source plane
magnification as a function of deflector--source impact parameter is
calculated for the deflector sample using a ray--tracing algorithm,
incorporating variations in deflector (e.g. central velocity
dispersion and redshift) and source (e.g. surface brightness
morphology and redshift) properties. To reproduce the effects of
atmospheric seeing, lensed images are convolved with a Gaussian kernel
of ${\rm{FWHM}} =  1 \scnd 5$, while to reproduce the 2dF
observations, the lens model considers the total flux received from a
$1 \scnp$ radius optical fibre centred on each deflector.

The sample of deflectors presents a magnified view of the distant
source population, characterised by a Lyman--$\alpha$ emission--line
luminosity function. The probability of detecting a lensed emission
line of given observed frame properties (i.e. flux, wavelength and
FWHM) drawn from this population is calculated via a `monte--carlo'
procedure whereby a grid of simulated emission lines of specified
properties are superimposed onto observed early--type spectra, to be
processed using the automated line detection algorithm.

\begin{figure}
\vbox{
\centerline{
\psfig{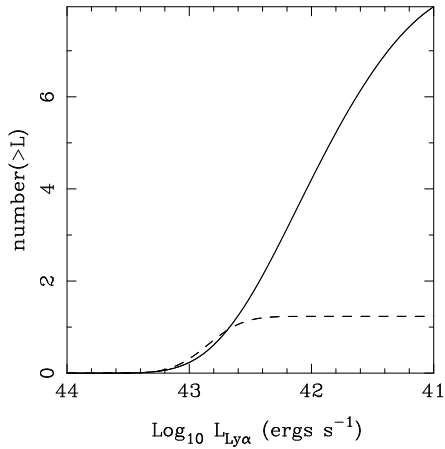}
}
\caption{Cumulative number distribution of gravitational lenses versus
intrinsic Lyman--$\alpha$ emission luminosity identified within a
projected sample of 2000 early--type galaxy spectra. Two model
luminosity functions are considered; a Schechter function described by
the parameters $L^{\ast} = 1 \times 10^{43}$ h$^2$ ergs s$^{-1}$,
$\alpha = -1.6$, $\phi^{\ast} = 1 \times 10^{-3}$ h$^3$ Mpc$^{-3}$
${(\log L )}^{-1}$ (solid line) and a Gaussian function of equal
$L^{\ast}$ and $\phi^{\ast}$ with a width parameter $\sigma = 0.25
\log L$ (dashed line). The model was realised using a cosmological
model specified by the parameters $\Omega=0.3$, $\Lambda=0.7$.}
\label{fig-2}
}
\end{figure}

Combining the magnification profiles generated by the deflector
sample with an assumed Lyman--$\alpha$ emission--line luminosity
function describing the source population, produces the number
distribution of detected lenses as a function of Lyman--$\alpha$
luminosity (Figure \ref{fig-2}). The number--luminosity diagram of
identified lenses generated by a spectroscopic sample of 2000
early--type galaxies, considering two competing luminosity function
models, clearly demonstrates the potential of the survey technique to
probe the characteristics of the faint Lyman--$\alpha$ emitting galaxy
population.

\section{Conclusions}

We have presented the strategy and initial observations for a
spectroscopic survey for strong galaxy--galaxy lenses. The optimal
line--of--sight strategy offers a powerful probe of the faint,
Lyman--$\alpha$ emitting, galaxy population and of the dark matter
profiles of massive early--type galaxies at cosmological
distances. Application of the deflector--based survey strategy to
other galaxy samples is underway \cite{hall00} and we hope to compile
a much larger sample of systems using additional 2dF observations.

\section*{Acknowledgments}

J. Willis acknowledges the support of a PPARC research studentship during
the completion of this work and the generous financial support of the
IoA, Cambridge and the European TMR network in order to attend the
conference.

\end{document}